\documentclass{svproc}
\usepackage{graphicx}%
\usepackage{multirow}%
\usepackage{amsmath,amssymb,amsfonts}%
\usepackage{mathrsfs}%
\usepackage[title]{appendix}%
\usepackage{xcolor}%
\usepackage{textcomp}%
\usepackage{manyfoot}%
\usepackage{booktabs}%
\usepackage{algorithm}%
\usepackage{algorithmicx}%
\usepackage{algpseudocode}%
\usepackage{listings}%
\usepackage{hyperref}
\usepackage{url}

\def\orcidID#1{\unskip$^{[#1]}$} % added MR 2018-03-10

\usepackage{subcaption}

\begin{document}
\mainmatter              % start of a contribution
\title{Textual understanding boost in the WikiRace}
\titlerunning{Textual understanding boost in the WikiRace}  % abbreviated title (for running head)
\author{Raman Ebrahimi\inst{1}\orcidID{0009-0004-6724-6029} \and Sean Fuhrman\inst{1}
Kendrick Nguyen\inst{1} \and Harini Gurusankar\inst{1} \and Massimo Franceschetti\inst{1}}
\authorrunning{Raman Ebrahimi et al.} % abbreviated author list (for running head)

\institute{University of California, San Diego, Electrical and Computer Engineering,\\
\email{raman@ucsd.edu}
California, USA}

\maketitle              % typeset the title of the contribution

\begin{abstract}
The WikiRace game, where players navigate between Wikipedia articles using only hyperlinks, serves as a compelling benchmark for goal-directed search in complex information networks. This paper presents a systematic evaluation of navigation strategies for this task, comparing agents guided by graph-theoretic structure (betweenness centrality), semantic meaning (language model embeddings), and hybrid approaches. Through rigorous benchmarking on a large Wikipedia subgraph, we demonstrate that a purely greedy agent guided by the semantic similarity of article titles is overwhelmingly effective. This strategy, when combined with a simple loop-avoidance mechanism, achieved a perfect success rate and navigated the network with an efficiency an order of magnitude better than structural or hybrid methods. Our findings highlight the critical limitations of purely structural heuristics for goal-directed search and underscore the transformative potential of large language models to act as powerful, zero-shot semantic navigators in complex information spaces. \footnote{The code used for the experiments in this paper is publicly available on \href{https://github.com/RamanEbrahimi/Research/tree/main/WikiRace-code}{Github}.}
\end{abstract}
\section{Introduction}

The challenge of navigating large-scale information networks to find a specific target is a fundamental problem in computer science and information retrieval. The WikiRace game, also known as Wikispeedia, provides a compelling and tractable research framework for studying this problem. In this game, a player is given a starting and a target Wikipedia article and must navigate from the former to the latter exclusively by clicking on hyperlinks embedded within the articles. The primary objective is typically to find a path with the minimum number of clicks, which frames the task as a shortest-path problem on the vast and complex graph formed by Wikipedia's hyperlink structure. This setup abstracts the real-world challenge of goal-directed search in an environment where global information about the network's structure is unavailable, and navigational decisions must be made based on local cues and semantic context \cite{west2015mining}.

The formalization of this game for academic research was pioneered by \cite{west2009wikispeedia}, who developed the Wikispeedia platform and the associated public dataset. This dataset has become a cornerstone for research in this domain. It consists of a condensed but substantial snapshot of Wikipedia, comprising 4,604 articles (nodes) and 119,882 hyperlinks (directed edges), along with a corpus of over 51,000 completed human navigation paths \cite{west2009wikispeedia,west2012human}. The availability of both the static graph structure and the dynamic, behavioral data from human gameplay has catalyzed a rich body of research into human navigation patterns, semantic distance, and algorithmic search strategies \cite{west2015mining}. The underlying graph exhibits properties common to real-world networks, such as a small-world structure and sparse connectivity, making it a realistic and challenging benchmark for developing and evaluating navigation algorithms.
The literature approaches the WikiRace problem from two distinct perspectives, creating a fundamental duality in the problem space. On one hand, it is treated as a purely graph-theoretic optimization problem, where the goal is to find the mathematically shortest path between two nodes, a task solvable by classical algorithms like Breadth-First Search (BFS). On the other hand, a significant body of work treats the game as a means to understand and model human ``common sense'' navigation \cite{west2009wikispeedia}. This research highlights that human-generated paths are systematically and predictably longer than the optimal shortest paths. This discrepancy arises because human players, lacking a global view of the graph, cannot execute an exhaustive search. Instead, they rely on local semantic cues, background knowledge, and cognitive heuristics to make their next move. Indeed, the Wikispeedia game was designed precisely to capture and quantify this latent common-sense knowledge embedded in human navigational choices. This duality establishes two different optimization targets for an artificial agent: achieving mathematical optimality by finding the shortest path, versus achieving cognitive plausibility by finding a human-like path. Any proposed navigation strategy must therefore be evaluated against one or both of these benchmarks, acknowledging that they represent distinct and often conflicting objectives.

\begin{figure}[ht]
    \includegraphics[width=\textwidth]{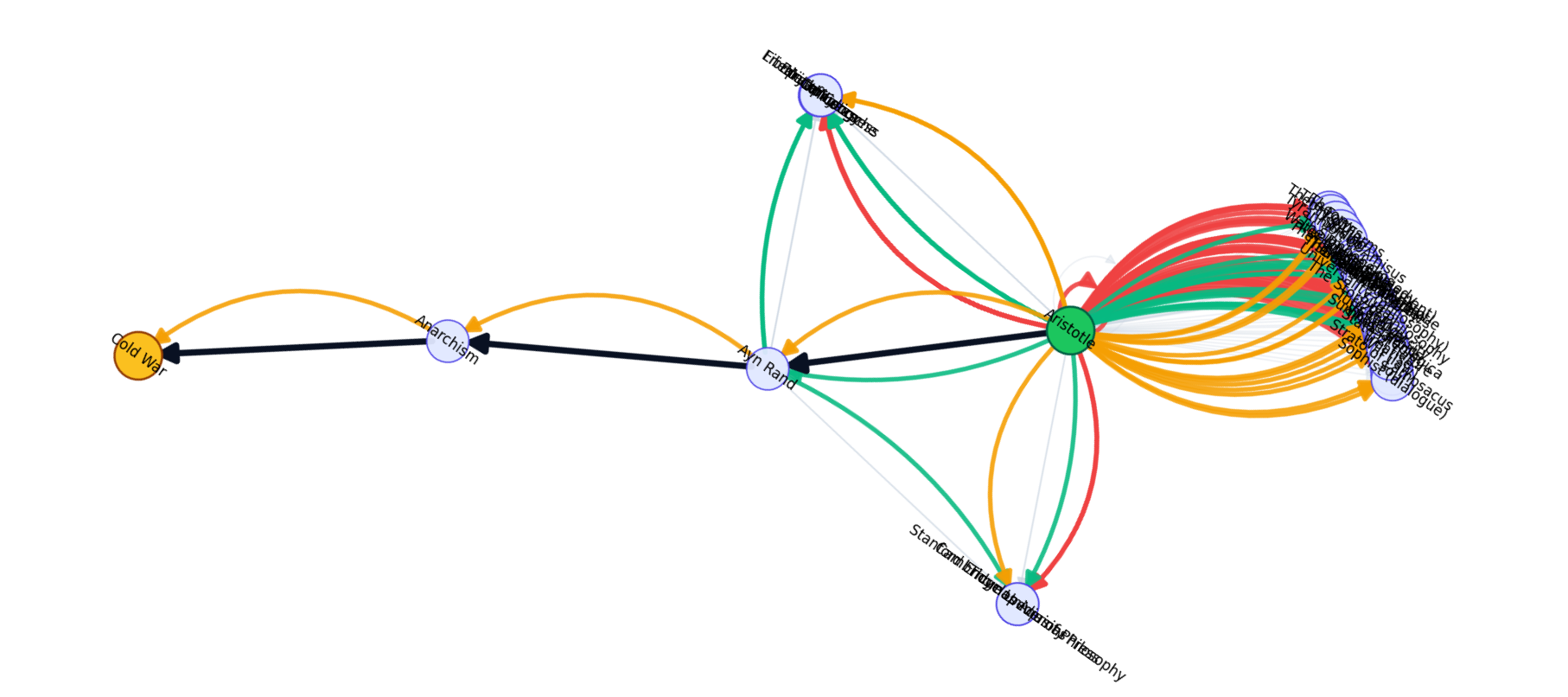}
    \caption{An instance of WikiRace on a small 1000 nodes subgraph starting from the node Aristotle (green) with the target node Cold War (yellow). Displayed strategies are random (red), $LLM^*$ (yellow), and betweenness (green). The optimal path is shown by the black links.}
    \vspace{-0.1in}
    \label{fig:instance}
\end{figure}

This paper presents a systematic investigation into various strategies for solving the WikiRace problem, as seen in Figure~\ref{fig:instance}, with a focus on comparing graph-theoretic heuristics against LLM-guided agents. Our contributions are threefold:
\begin{enumerate}
    \item We design, formalize, and implement a diverse set of navigational strategies. These include a random walk baseline, a heuristic based on betweenness centrality, several LLM-driven agents with varying levels of contextual awareness, and hybrid models that combine structural and semantic approaches.
    \item We introduce a novel context-enhanced LLM agent that is prompted with rich, local semantic information from the current article. We further augment this agent with an $\epsilon$-greedy action selection policy, a technique borrowed from reinforcement learning, to systematically manage the trade-off between exploiting the LLM's best judgment and exploring alternative paths.
    \item Through rigorous empirical evaluation, we demonstrate that a context-aware, purely greedy LLM agent achieves near-perfect performance, significantly outperforming all other strategies. This finding challenges conventional assumptions about the necessity of exploration in heuristic search and provides critical evidence for the efficacy of LLMs as powerful, zero-shot semantic navigators.
\end{enumerate}

\section{Related Work}
The analysis of human navigation paths in the Wikispeedia dataset has revealed consistent cognitive strategies that serve as a crucial benchmark for algorithmic approaches. These human strategies are not random but are guided by sophisticated, albeit imperfect, heuristics for managing the complexity of the information space.

Goal-directed navigation often follows a two-phase ``zoom-out, home-in'' heuristic \cite{siegart-2020}. Initially, players navigate from specific starting articles to broader, central ``hubs'' to escape isolated regions of the network \cite{west2009wikispeedia}. After reaching a hub, they enter the ``homing-in'' phase, selecting links semantically closer to the target \cite{lamprecht2017structure}. This contrasts with free-form browsing, which typically follows a path of increasing specificity \cite{rodi2017search}.

The ``zoom-out'' phase involves a structural leap, where the first click leads to an article with a much higher degree \cite{west2012human}. The subsequent ``homing-in'' phase is guided by semantic coherence, as the textual similarity between the current and target articles steadily increases \cite{west2012human}. This indicates that humans use a decentralized search strategy based on an implicit sense of semantic distance \cite{rodi2017search}. While shortest-path algorithms also use high-degree nodes, they do so as a byproduct of high betweenness centrality, not as a deliberate two-phase cognitive strategy.

This heuristic makes human paths less efficient than the mathematical optimum, often involving ``circling'' behavior near the target \cite{west2009wikispeedia,west2012human}. However, this apparent inefficiency is a feature of a robust cognitive strategy for navigating with incomplete information. The ``zoom-out, home-in'' heuristic manages uncertainty by moving to a well-connected hub, which increases the probability of finding a route to the target's semantic neighborhood. This strategy prioritizes success over optimal path length. Therefore, human ``inefficiency'' is a key trait of effective search under constraints, suggesting navigation agents should be evaluated on their ability to model this trade-off, not just on path length.

\subsection{Graph-theoretic Heuristics: Centrality-based Navigation}
Graph-theoretic heuristics guide WikiRace navigation by using centrality measures to quantify the importance of an article within the network's structure. Common measures like \emph{Degree Centrality}, \emph{Closeness Centrality}, \emph{Betweenness Centrality}, and \emph{PageRank} are used to identify nodes that are local hubs, efficient information spreaders, critical bridges between topics, or general authorities, respectively \cite{koschutzki2008centrality,cheriyan2020m}. A key distinction lies between radial measures (Degree, PageRank), which identify destination hubs, and medial measures (Betweenness), which identify intermediary bridges \cite{borgatti2006graph}.

However, these heuristics face a fundamental paradox in goal-directed navigation. While maximizing centrality effectively supports the human "zoom-out" strategy of moving from a specific article to a major hub, it inhibits the subsequent ``home-in'' phase. A purely centrality-based approach becomes trapped at high-importance nodes, as it cannot justify a move to a less central but more semantically relevant article closer to the target. This highlights the limitation of using static structural importance for a multi-stage navigation task.

\subsection{Semantic and Learning-based Navigation Strategies}
To overcome the limitations of purely structural heuristics, advanced strategies incorporate the semantic content of articles using Natural Language Processing (NLP) and machine learning. These methods model the ``homing-in'' phase of navigation by selecting links that are most semantically similar to the target.

The primary challenge is effectively representing article meaning. Methods have evolved from statistical models like Term Frequency-Inverse Document Frequency (TF-IDF) to more powerful dense vector embeddings. Models such as Word2Vec \cite{church2017word2vec} and the specialized Wikipedia2Vec \cite{yamada2018wikipedia2vec} learn vectors that capture nuanced semantic relationships, providing a more robust measure of similarity than sparse lexical methods \cite{kwon2020hierarchical}.

A parallel line of research frames WikiRace as a formal learning problem. One approach learns a heuristic function, $h(v)$, to predict the shortest path distance from a given article $v$ to the target. This learned heuristic can then be integrated into a classical search algorithm like $A^*$ to guide exploration more efficiently \cite{barron-2011}. Features for such models range from bag-of-words vectors to article metadata or cluster IDs \cite{barron-2011}.

The problem can also be modeled for Reinforcement Learning (RL) \cite{darvariu2024graph}, where an agent learns a policy, $\pi(a|s)$, to select a hyperlink that maximizes a cumulative reward. Given the large corpus of human gameplay data, Imitation Learning (IL) is a particularly suitable framework. For instance, Behavioral Cloning can train a policy to mimic human actions. In a significant result, \cite{zaheer2022learning} demonstrated that a policy learned via behavioral cloning on randomly sampled trajectories is sufficient to create an effective navigation agent for the full Wikipedia graph.

The evolution of these strategies reflects a broader trend from systems based on hand-crafted heuristics toward end-to-end learning. While modern models achieve impressive performance, their decision-making can be opaque. This creates a compelling research opportunity for developing strategies that are both high-performing and interpretable, potentially by combining the structural logic of network science with the representation power of modern machine learning.

A clear gap exists at the intersection of these approaches. On one side are purely structural methods like centrality, which are interpretable but too simplistic to capture the full complexity of the task. On the other are purely semantic or black-box learning methods, which are powerful but may ignore the fundamental network properties that guide human intuition and are difficult to analyze. This points to the need for novel network-based strategies that explicitly and intelligently integrate both the topological structure of the graph and the semantic content of its nodes. Such hybrid methods could potentially capture the best of both worlds: the structural awareness of centrality and the semantic nuance of modern NLP, leading to navigation policies that are not only more effective but also more interpretable and human-like. Developing such strategies represents a promising direction for future research.

\section{Methodology}
Our methodology is divided into three stages: (1) constructing a tractable, high-quality subgraph from the comprehensive Wikipedia link graph; (2) formalizing and implementing a diverse set of navigation strategies, ranging from simple baselines to sophisticated semantic and hybrid agents; and (3) establishing a rigorous benchmarking framework to evaluate the performance of these strategies.

\subsection{Data Preparation and Graph Construction}
We began with the publicly available WikiLinkGraphs dataset \cite{consonni2019wikilinkgraphs}, specifically the \texttt{enwiki.wikilink\_graph.2018-03-01} snapshot. This dataset represents Wikipedia as a directed graph where articles are nodes and hyperlinks are edges. Given its massive scale (over 5 million nodes and 100 million edges), we developed a multi-step pipeline to create a smaller, yet representative, subgraph suitable for iterative experimentation.

\textbf{Initial Ingestion and Pruning:} We first processed the full dataset using a streaming approach to manage memory constraints. The gzipped TSV file was read in chunks. During this process, we built a directed graph using the NetworKit library \cite{staudt2016networkit}, which is highly optimized for large-scale network analysis. Concurrently, we extracted a mapping of unique node IDs to their corresponding article titles, which was serialized to an efficient Parquet file format for later use by semantic strategies. Following the initial construction, we performed a critical pruning step, iteratively removing all sink nodes (nodes with an out-degree of 0). This process ensures that every node in the resulting core graph has at least one incoming and one outgoing link, eliminating dead ends and creating a more strongly connected component for navigation.

\textbf{Subgraph Extraction:} To create a final, manageable graph for our experiments, we employed a $k$-ball sampling method. We selected a random seed node from the pruned core graph and performed a Breadth-First Search (BFS) up to a specified radius $k$. The resulting set of nodes within this $k$-hop neighborhood forms our subgraph. This approach preserves the local connectivity and community structure around the seed node. For our experiments, we set a target size of approximately 100,000 nodes and a radius of $k=3$. The final subgraph used for benchmarking consisted of all nodes within this ball and the induced edges between them from the original graph. This subgraph was saved in both the NetworKit binary format for fast loading and the standard GraphML format for compatibility.

\subsection{Navigation Strategies}
Let the Wikipedia graph be a directed graph $G=(V,E)$, where $V$ is the set of articles and $E$ is the set of hyperlinks. The WikiRace task is to find a path from a starting node $v_s \in V$ to a target node $v_g\in V$. At any step $t$, the agent is at a current node $v_c$ and must select a successor node $v_\text{next}\in N^+(v_c)$, where $N^+(v_c)=\{v'|(v_c, v')\in E\}$ is the set of nodes linked from $v_c$. We designed and implemented the following strategies to select $v_\text{next}$.

\textbf{Baselines:} To contextualize performance, we included three crucial benchmarks: (1) \textit{Random:} A baseline that randomly selects a neighbor at each step. (2) \textit{Human Expert:} A human expert (a friend of the authors) played the same 10 test games given to the algorithms, providing a benchmark for human-level performance. (3) \textit{Shortest-Path:} An oracle that computes the mathematically optimal shortest path, representing the theoretical best possible score. The random strategy, serves as a simple baseline. At each step, the next node is chosen uniformly at random from the set of successors: $v_\text{next} = \textbf{UniformChoice}(N^+(v_c))$. This strategy ignores the goal and relies purely on the graph's topology and chance.

\textbf{Betweenness Centrality Heuristic:} This strategy, uses a structural heuristic to guide navigation. It prioritizes moving to nodes that are more central to the graph's structure, under the assumption that such nodes are important crossroads. We use betweenness centrality, $C_B(v)$, which measures the fraction of all-pairs shortest paths that pass through node $v$. At each step, the agent selects the neighbor with the highest pre-computed betweenness centrality: $v_\text{next} = \arg\max_{v'\in N^+(v_c)} C_B(v')$. To make the computation of $C_B(v)$ tractable on our large subgraph, we use a stochastic approximation calculated on a sample of $k=512$ nodes. The implementation also includes a simple loop avoidance mechanism by preventing immediate backward moves.

\textbf{LLM-based Semantic Navigation:} These strategies leverage the semantic understanding of a pre-trained language model to guide the search. We use the \texttt{all-MiniLM-L6-v2} sentence-transformer model to generate a 384-dimensional embedding vector $\Phi(v)$ for the title of each article $v$. The agent's policy is to select the neighbor whose title is most semantically similar to the target's title, as measured by the cosine similarity of their embeddings.

\textbf{Greedy Semantic Search:} This is a purely greedy strategy.
\begin{align}
    v_\text{next} = \arg\max_{v'\in N^+(v_c)}\frac{\Phi(v')\dot \Phi(v_g)}{||\Phi(v')||\;||\Phi(v_g)||}
\end{align} 
The embeddings are pre-normalized, so this simplifies to a dot product. Node embeddings are cached upon first computation for efficiency.

\textbf{Greedy Semantic Search with Loop Avoidance (\texttt{llm-extra-eps-0}):} This variant enhances the greedy approach by maintaining a set of visited nodes, $\mathcal{V}_\text{path}$, for the current game. It prioritizes selecting from the set of unvisited neighbors, $N^+(v_c)\backslash\mathcal{V}_{path}$, falling back to the full neighbor set only if all neighbors have been visited. The \texttt{eps-0} signifies a purely greedy selection policy.

\subsection{Hybrid Strategies}
We designed two hybrid strategies that combine structural and semantic information, attempting to mimic the two-phase ``zoom-out, home-in'' heuristic observed in human navigation.

\textbf{Betweenness-then-LLM (betweenness-then-llm):} This strategy follows a fixed two-phase approach. For the first three hops of any game, it uses the betweenness centrality heuristic to ``zoom out'' to a central part of the graph. After the third hop, it switches permanently to the greedy LLM semantic strategy to ``home in'' on the target.

\textbf{LLM with Betweenness Fallback (llm-fallback):} This strategy dynamically adapts its approach at each step. It primarily relies on the greedy LLM semantic strategy. However, if the maximum similarity score among all neighbors falls below a predefined threshold $\theta=0.25$, it deems the semantic signal to be too weak. In such cases, it ``falls back'' to using the betweenness centrality heuristic for that single step. Let $\mathcal{S} = \textbf{sim}(\Phi(v'), \Phi(v_g))$, then:
\[ v_\text{next} = \begin{cases} 
      \arg\max_{v'\in N^+(v_c)} \mathcal{S} & \text{if } \max_{v'\in N^+(v_c)} \mathcal{S} \ge \theta\\
      \arg\max_{v'\in N^+(v_c)} C_B(v') & \text{otherwise}
   \end{cases}
\] 
Based on initial observations that simple greedy strategies often get stuck in cycles, we introduced several variations. (1) \textit{Loop Prevention (* suffix):} Strategies marked with an asterisk (e.g., Betweenness*, LLM*) are enhanced with a memory of visited nodes. At each step, the agent is restricted to choosing from unvisited neighbors preventing loops. (2) \textit{Epsilon-Greedy Exploration (LLM* + eps):} This strategy introduces exploration. With a probability of $\epsilon=0.1$, it selects a random neighbor; otherwise (with probability $1-\epsilon$), it follows the greedy LLM* policy. (3) \textit{LLM Size Variation (LLM-L, LLM-XL):} To test the impact of model scale, we implemented LLM* with larger and extra-large language models.

\label{sec:experiments}
\section{Experiments and Results}

\subsubsection{Experimental Setup}
We evaluated all strategies on the same induced subgraph of the Wikipedia network. For each strategy, we conducted a benchmark of $n=10$ WikiRace games. To ensure reproducibility and fair comparison, all runs used a fixed random seed (seed=42). This seed determined the selection of goal nodes for the 10 games, ensuring every strategy was tested on the same set of tasks. A universal, fixed starting node was used for all games. We set a maximum path length of 5,000 hops; if an agent failed to reach the target within this limit, the game was recorded as a failure with a hop count of 5,000. The primary metric for evaluation is the average number of hops taken to reach the goal across the 10 trials.

\subsubsection{Results}
The performance of each navigation strategy is summarized in Table 1. The results reveal a stark contrast between the effectiveness of structural, semantic, and hybrid approaches.
\begin{table}[t]
\label{tab:results}
\caption{Performance of WikiRace navigation strategies. Results are averaged over 10 test runs, with a failure cap of 5,000 hops.}
\begin{center}
\begin{tabular}{lll}
\multicolumn{1}{c}{\bf Strategy}  &\multicolumn{1}{c}{\bf Average Hops} &\multicolumn{1}{c}{\bf Success Rate}
\\ \hline \\
\textit{Oracle \& Human Benchmark} & & \\
Shortest-Path & 12.4 & 100\% \\
Human Expert  & 45.1 & 100\% \\
\hline \\
\textit{Baseline} & & \\
Random  & 958.2 & 90\% \\
Betweenness  & 5000.0 & 0\% \\
LLM  & 3502.1 & 30\% \\
\hline \\
\textit{Core Strategies with Enhancements} & & \\
Betweenness*  & 4055.6 & 20\% \\
\textbf{LLM*}  & \textbf{155.3} & \textbf{100\%} \\
LLM*+eps  & 217.9 & 100\% \\
\hline \\
\textit{LLM Variations} & & \\
LLM-L (Large)  & 189.4 & 100\% \\
LLM-XL (Extra Large)  & 201.5 & 100\% \\
\hline \\
\textit{Hybrid Strategies} & & \\
Betweenness+LLM* (K=3)  & 2004.2 & 50\% \\
LLM* Fallback  & 1572.8 & 70\% \\
\end{tabular}
\end{center}
\end{table}

The most striking result is the difference between strategies with and without the * modifier. The basic Betweenness and LLM agents failed 100\% and 70\% of the time, respectively, by hitting the 5,000 hop limit. This demonstrates that getting trapped in cycles is a primary failure mode for greedy navigation, and that implementing a simple visited-set memory is crucial for success. The Betweenness* strategy performed very poorly, even worse than the Random baseline. We hypothesize this is because betweenness centrality guides the agent toward general-purpose hubs, not necessarily nodes that are semantically closer to the specific goal. The goal article is not guaranteed to be near a high-centrality node, making this heuristic an unreliable compass. This result confirms the limitation of purely structural heuristics: while effective at identifying global hubs, they are unable to navigate towards specific, less central targets. The LLM-based strategies demonstrated vastly superior performance. The poor performance of the hybrid models (Betweenness+LLM* and LLM* Fallback), which failed to improve upon LLM*, further suggests that betweenness centrality is not a useful signal for this task. The hybrid strategies, designed to balance structural and semantic guidance, performance was severely hampered by games where the initial structural guidance led the agent astray, from which the semantic component could not recover within the hop limit. This suggests that the ``structural zoom-out'' phase, when guided by a simplistic centrality metric, may be more detrimental than helpful, and that the LLM's semantic guidance is robust enough on its own.

The LLM* agent was the clear winner among the automated strategies. Its success highlights the power of semantic similarity for navigation. Interestingly, using larger LLM variants (-L, -XL) slightly degraded performance. While this may be due to variance from the small number of test runs, it suggests that for encoding short phrases like article titles, increasing model size beyond a certain point may not be beneficial. This highlights the profound effectiveness of using semantic similarity as the primary navigational compass. The guidance provided by the language model is so precise that a purely greedy approach is sufficient to navigate the complex graph with remarkable efficiency. 

\section{Discussion}
Our experimental results provide a clear and compelling conclusion: for the WikiRace navigation task, a purely greedy semantic search guided by language model embeddings is overwhelmingly effective. The success of the LLM* strategy, which combines a strong semantic compass with basic loop avoidance, challenges several assumptions in heuristic search.

First, it suggests that in semantically rich networks like Wikipedia, the need for exploration (e.g., an $\epsilon$-greedy policy) may be minimal when a sufficiently powerful heuristic is available. The LLM's ability to consistently identify the most promising neighbor creates a ``semantic gradient'' so steep that deviating from it is rarely optimal. The agent effectively surfs this gradient directly to the target. 

Second, our findings highlight the critical limitations of purely structural heuristics like betweenness centrality for goal-directed navigation. While central nodes are vital to the graph's overall connectivity, a strategy that perpetually seeks them becomes trapped in a ``centrality gravity well,'' unable to make progress towards a specific, and likely less central, target. This confirms the paradox discussed in our related work: the optimal strategy for the ``zoom-out'' phase is counterproductive for the ``home-in'' phase.

Finally, the underwhelming performance of our hybrid models suggests that a naive combination of structural and semantic signals is insufficient. The betweenness-then-llm model, in particular, was an explicit attempt to codify the human ``zoom-out, home-in'' strategy. Its failure implies that human navigation is likely guided by a more nuanced, context-aware integration of topology and semantics than our simple, two-stage model can capture.

\section{Conclusion}
In this paper, we presented a systematic evaluation of graph-theoretic, semantic, and hybrid strategies for solving the WikiRace game. We developed a robust data pipeline to construct a representative subgraph of Wikipedia and implemented a diverse suite of navigation agents. Our rigorous benchmarking demonstrates that a greedy agent guided by the semantic similarity of article titles, as encoded by a sentence-transformer model, dramatically outperforms all other strategies. This simple yet powerful approach achieved a perfect success rate, navigating the complex information network with an efficiency an order of magnitude greater than structural or hybrid methods. Our findings underscore the transformative potential of large language models as zero-shot navigators in complex information spaces and establish a new, formidable benchmark for future research in goal-directed search on networks.

Our study was conducted on a subgraph of Wikipedia. While this was necessary for computational tractability, future work should validate these findings on the full, unabridged graph. Furthermore, our semantic strategies relied solely on article titles. Incorporating the full text of articles or abstracts could provide a richer semantic signal and potentially lead to even shorter paths. A promising direction for hybrid models would be to move beyond simple centrality and explore Graph Neural Network (GNN) architectures that can learn to integrate local graph structure and node features (text embeddings) into a unified navigational policy.

% ---- OR ------- Uncomment this section to use bibtex
\bibliographystyle{spmpsci} % We choose the "plain" reference style
\bibliography{refs} % Entries are in the refs.bib file

@inproceedings{west2015mining,
  title={Mining missing hyperlinks from human navigation traces: A case study of Wikipedia},
  author={West, Robert and Paranjape, Ashwin and Leskovec, Jure},
  booktitle={Proceedings of the 24th international conference on World Wide Web},
  pages={1242--1252},
  year={2015}
}

@inproceedings{west2009wikispeedia,
  title={Wikispeedia: An Online Game for Inferring Semantic Distances between Concepts.},
  author={West, Robert and Pineau, Joelle and Precup, Doina},
  booktitle={IJCAI},
  volume={9},
  pages={1598--1603},
  year={2009}
}

@inproceedings{west2012human,
  title={Human wayfinding in information networks},
  author={West, Robert and Leskovec, Jure},
  booktitle={Proceedings of the 21st international conference on World Wide Web},
  pages={619--628},
  year={2012}
}

@misc{siegart-2020,
	author = {Siegart, Jamie},
	month = {4},
	title = {{A Network Analysis of Wikispeedia}},
	year = {2020},
	url = {https://jsiegart.medium.com/a-network-analysis-of-wikispeedia-da4525e1f207},
}

@article{lamprecht2017structure,
  title={How the structure of Wikipedia articles influences user navigation},
  author={Lamprecht, Daniel and Lerman, Kristina and Helic, Denis and Strohmaier, Markus},
  journal={New Review of Hypermedia and Multimedia},
  volume={23},
  number={1},
  pages={29--50},
  year={2017},
  publisher={Taylor \& Francis}
}

@article{rodi2017search,
  title={Search strategies of Wikipedia readers},
  author={Rodi, Giovanna Chiara and Loreto, Vittorio and Tria, Francesca},
  journal={PloS one},
  volume={12},
  number={2},
  pages={e0170746},
  year={2017},
  publisher={Public Library of Science San Francisco, CA USA}
}

@article{koschutzki2008centrality,
  title={Centrality analysis methods for biological networks and their application to gene regulatory networks},
  author={Kosch{\"u}tzki, Dirk and Schreiber, Falk},
  journal={Gene regulation and systems biology},
  volume={2},
  pages={GRSB--S702},
  year={2008},
  publisher={SAGE Publications Sage UK: London, England}
}

@article{cheriyan2020m,
  title={m-PageRank: A novel centrality measure for multilayer networks},
  author={Cheriyan, Jo and Sajeev, GP},
  journal={Advances in Complex Systems},
  volume={23},
  number={05},
  pages={2050012},
  year={2020},
  publisher={World Scientific}
}

@article{borgatti2006graph,
  title={A graph-theoretic perspective on centrality},
  author={Borgatti, Stephen P and Everett, Martin G},
  journal={Social networks},
  volume={28},
  number={4},
  pages={466--484},
  year={2006},
  publisher={Elsevier}
}

@article{church2017word2vec,
  title={Word2Vec},
  author={Church, Kenneth Ward},
  journal={Natural Language Engineering},
  volume={23},
  number={1},
  pages={155--162},
  year={2017},
  publisher={Cambridge University Press}
}

@article{yamada2018wikipedia2vec,
  title={Wikipedia2Vec: An efficient toolkit for learning and visualizing the embeddings of words and entities from Wikipedia},
  author={Yamada, Ikuya and Asai, Akari and Sakuma, Jin and Shindo, Hiroyuki and Takeda, Hideaki and Takefuji, Yoshiyasu and Matsumoto, Yuji},
  journal={arXiv preprint arXiv:1812.06280},
  year={2018}
}

@inproceedings{kwon2020hierarchical,
  title={Hierarchical trivia fact extraction from Wikipedia articles},
  author={Kwon, Jingun and Kamigaito, Hidetaka and Song, Young-In and Okumura, Manabu},
  booktitle={Proceedings of the 28th International Conference on Computational Linguistics},
  pages={4825--4834},
  year={2020}
}

@techreport{barron-2011,
	author = {Barron, Alex and Swafford, Zack},
	title = {{An AI for the Wikipedia game}},
	year = {2011},
	url = {https://cs229.stanford.edu/proj2015/309_report.pdf},
}

@article{darvariu2024graph,
  title={Graph reinforcement learning for combinatorial optimization: A survey and unifying perspective},
  author={Darvariu, Victor-Alexandru and Hailes, Stephen and Musolesi, Mirco},
  journal={arXiv preprint arXiv:2404.06492},
  year={2024}
}

@article{zaheer2022learning,
  title={Learning to navigate wikipedia by taking random walks},
  author={Zaheer, Manzil and Marino, Kenneth and Grathwohl, Will and Schultz, John and Shang, Wendy and Babayan, Sheila and Ahuja, Arun and Dasgupta, Ishita and Kaeser-Chen, Christine and Fergus, Rob},
  journal={Advances in Neural Information Processing Systems},
  volume={35},
  pages={1529--1541},
  year={2022}
}

@inproceedings{consonni2019wikilinkgraphs,
  title={WikiLinkGraphs: A complete, longitudinal and multi-language dataset of the Wikipedia link networks},
  author={Consonni, Cristian and Laniado, David and Montresor, Alberto},
  booktitle={Proceedings of the International AAAI Conference on Web and Social Media},
  volume={13},
  pages={598--607},
  year={2019}
}

@article{staudt2016networkit,
  title={NetworKit: A tool suite for large-scale complex network analysis},
  author={Staudt, Christian L and Sazonovs, Aleksejs and Meyerhenke, Henning},
  journal={Network Science},
  volume={4},
  number={4},
  pages={508--530},
  year={2016}
}

\end{document}